\documentclass[12pt]{article}
\usepackage[dvips]{graphicx}
\usepackage{epsfig,cite}
\usepackage {amsmath}
\usepackage{color}
\usepackage{amssymb}
\usepackage{slashed}
\include{epsf} 
\newcount\timecount
\newcount\hours \newcount\minutes  \newcount\temp \newcount\pmhours
\hours = \time
\divide\hours by 60
\temp = \hours
\multiply\temp by 60
\minutes = \time
\advance\minutes by -\temp
\def\hour{\the\hours}
\def\minute{\ifnum\minutes<10 0\the\minutes
            \else\the\minutes\fi}
\def\clock{
\ifnum\hours=0 12:\minute\ AM
\else\ifnum\hours<12 \hour:\minute\ AM
      \else\ifnum\hours=12 12:\minute\ PM
            \else\ifnum\hours>12
                 \pmhours=\hours
                 \advance\pmhours by -12
                 \the\pmhours:\minute\ PM
                 \fi
            \fi
      \fi
\fi
}

\def\monthname{\relax\ifcase\month 0/\or January\or February\or
   March\or April\or May\or June\or July\or August\or September\or
   October\or November\or December\else\number\month/\fi}

\def\bold#1{\setbox0=\hbox{$#1$}%
     \kern-.025em\copy0\kern-\wd0
     \kern.05em\copy0\kern-\wd0
     \kern-.025em\raise.0433em\box0 }

\newcommand{\be}{\begin{equation}}
\newcommand{\ee}{\end{equation}}
\newcommand{\bear}{\begin{eqnarray}}
\newcommand{\eear}{\end{eqnarray}}

\newcommand{\vev}[1]{\left\langle #1\right\rangle}

\newcommand{\lapproxeq}{\lower .7ex\hbox{$\;\stackrel{\textstyle  
<}{\sim}\;$}} 
\newcommand{\gapproxeq}{\lower .7ex\hbox{$\;\stackrel{\textstyle  
>}{\sim}\;$}} 
\newcommand{\stackdown}[2]{\lower 1.4ex\hbox{$\;\stackrel{\textstyle{#1}}  
{\scriptstyle{#2}}\;$}}
\newcommand{\beq}{\begin{equation}} 
\newcommand{\eeq}{\end{equation}} 
\newcommand{\ba}{\begin{eqnarray}}
\newcommand{\ea}{\end{eqnarray}}

\newcommand{\bea}{\begin{eqnarray}}
\newcommand{\eea}{\end{eqnarray}}

\def\gappeq{\mathrel{\rlap {\raise.5ex\hbox{$>$}}
{\lower.5ex\hbox{$\sim$}}}}
\def\lappeq{\mathrel{\rlap{\raise.5ex\hbox{$<$}}
{\lower.5ex\hbox{$\sim$}}}}
\def\Toprel#1\over#2{\mathrel{\mathop{#2}\limits^{#1}}}

\makeatletter 
\def\slash{\@ifnextchar[{\fmsl@sh}{\fmsl@sh[0mu]}} 
\def\fmsl@sh[#1]#2{% 
  \mathchoice 
    {\@fmsl@sh\displaystyle{#1}{#2}}% 
    {\@fmsl@sh\textstyle{#1}{#2}}% 
    {\@fmsl@sh\scriptstyle{#1}{#2}}% 
    {\@fmsl@sh\scriptscriptstyle{#1}{#2}}} 
\def\@fmsl@sh#1#2#3{\m@th\ooalign{$\hfil#1\mkern#2/\hfil$\crcr$#1#3$}} 
\makeatother 
\begin{document} 
\begin{titlepage} 
 
%%%%%%%%%%% 
\begin{flushright}
%% \today \\
ACT-04-20 \\  
MI-TH-2015\\
%\today
 \end{flushright} 
%%%%%%%%%% 
\begin{center} 
\vspace*{1.2cm} 
 
{\Large{\textbf {{Primordial Black Holes  from No-Scale Supergravity}}}}\\ 
 \vspace*{10mm} 
 
{\bf  Dimitri~V.~Nanopoulos}$^1$,  {\bf  Vassilis~C.~Spanos}$^2$,  {\bf  Ioanna~D.~Stamou}$^2$ \\ 

\vspace{.7cm}

$^1${\em George P. and Cynthia W. Mitchell Institute for Fundamental
 Physics and Astronomy, Texas A\&M University, College Station, Texas 77843, USA;\\ 
 Astroparticle Physics Group, Houston Advanced Research Center (HARC),
 \\ Mitchell Campus, Woodlands, Texas 77381, USA;\\ 
Academy of Athens, Division of Natural Sciences,
Athens 10679, Greece}\\[0.3cm]

 $^2${\it National and Kapodistrian University of Athens, Department of Physics, \\
 Section of Nuclear {\rm \&} Particle Physics,  GR--15784 Athens, Greece}

\end{center} 
\vspace{2.3cm}
 
\begin{abstract}
We calculate the primordial black hole abundance in the context 
of a Wess-Zumino type no-scale supergravity model. We modify 
  the K\"ahler potential, by adding   an  extra exponential term. 
Using just one parameter  in the context of this model, we are  able to  satisfy  the Planck cosmological constraints for
the spectral index $n_s$, the tensor-to-scalar ratio $r$,  and  to  
produce  up to $\sim 20\%$ of the dark matter of the Universe in the form of  primordial black holes. 
\end{abstract}
\end{titlepage} 

\baselineskip=17pt

\section{Introduction}
\label{intro}

The recent observations of black hole (BH) mergers  by VIRGO/LIGO open a new window to probe BH
 physics \cite{Abbott:2016blz,Abbott:2016nmj,Abbott:2017vtc,Abbott:2017gyy,Abbott:2017oio}. 
 These detections rekindled the idea that Primordial Black Holes (PBH) can be considered as Dark Matter (DM) 
 candidates \cite{GarciaBellido:1996qt,Leach:2000yw,Green:2020jor}. As the nature of DM remains one of the most notable mysteries in physics, a  flurry of activity has recently  
 taken  place in this direction \cite{Ballesteros:2017fsr,Ozsoy:2018flq,Biagetti:2018pjj,Franciolini:2018vbk,Gao:2018pvq,Cicoli:2018asa,Dalianis:2018frf,Garcia-Bellido:2017mdw,Ezquiaga:2017fvi,Gong:2017qlj,Hertzberg:2017dkh,Espinosa:2017sgp,Clesse:2015wea,Germani:2017bcs,Inomata:2017okj,Motohashi:2017kbs,Kannike:2017bxn,Inomata:2017vxo,Kawasaki:2016pql,Carr:2016drx,Cheng:2018yyr,Mahbub:2019uhl,Mishra:2019pzq,Ballesteros:2019hus,Braglia:2020eai,Cai:2019bmk,Aldabergenov:2020bpt,Ketov:2019mfc}.

It has been proposed that a spike in the Cosmic Microwave  Background (CMB) power spectrum can be  physically significant,  as it could  lead  to  formation  of  PBHs.  
Such a  spike  is related to an inflection point in the scalar inflaton potential ~\cite{Garcia-Bellido:2017mdw}.
 In the context of   single field inflation models,  an  inflection point is  created  whence  the slow-roll parameter $\varepsilon$, that is   related to the derivative 
of the inflaton, gets sizeable  value. On the other hand,  $\varepsilon$ stays below one,  allowing the inflation to  goes on. The local enhancement supervened  
by a period where the inflaton is almost constant. During this plateau the power spectrum 
amplifies, enabling production of PBH in the   radiation dominated phase of the early universe. This PBH abundance  can be interpreted as a substantial fraction of the DM of the Universe. 
Similar reinforcement in the power spectrum can be 
achieved in the context of two-field models~\cite{Clesse:2015wea,Braglia:2020eai}. In these models, one field plays the role of the inflaton and the other is responsible 
for the PBH production.

It is now clear,  
 that a more precise calculation of  the power spectrum  is indispensable. 
This evaluation can be achieved by  solving numerically the so-called Mukhanov-Sasaki (M-S) equation \cite{Mukhanov:1988jd,Sasaki:1986hm}.
 Because the slow-roll approximation fails to reproduce the exact results in many proposed models, such as the one in ref.\cite{Ballesteros:2017fsr}, it is imperative to solve the M-S equation exactly. In addition to that, the precise  size and the location  of the peak of the power spectrum is crucial for calculating the fractional abundance of PBH in the Universe.

Here, we try to sum up  the basic developments in PBH production using  single field inflation. Specifically, in~\cite{Ballesteros:2017fsr} the authors employ a model based on  
an effective potential with an approximate inflection point, arise  from  two-loop logarithmic  corrections.
In~\cite{Ozsoy:2018flq,Cicoli:2018asa} it has been considered  the PBH production  studying a  string inflation  model.
Alternatively, models in (critical)  Higgs inflation has been studied in~ \cite{Ezquiaga:2017fvi,Espinosa:2017sgp}.  
A power spectrum by a polynomical potential has been suggested in~\cite{Garcia-Bellido:2017mdw,Hertzberg:2017dkh,Gong:2017qlj}. 
In~\cite{Gao:2018pvq,Kawasaki:2016pql} has been proposed a supergravity model with a single chiral field.
 Moreover, the authors in~\cite{Dalianis:2018frf,Mahbub:2019uhl} have studied   inflationary $\alpha$-attractor 
models.    Finally, PBH by axion monodromy has been considered in \cite{Ballesteros:2019hus}.

Embedding  models of inflation, into a more fundamental quantum theory such as supergravity, results to  a  framework that can be predictive
and reveals an aspect of the high energy scale~\cite{Ellis:1983sf}.    
 Taking this into account, we consider that  the natural framework for formulating models of inflation is supergravity.
 Specifically, no-scale supergravity models~\cite{Ellis:1983sf,Cremmer:1983bf,Ellis:1983ei,Ellis:1984bm,Lahanas:1986uc},  turn out to have other advantages:  their potential 
 depends on a minimal number  of parameters, they  evade the $\eta$ problem and they  emerge  naturally as the low energy limit of  compactified string models~\cite{Witten:1985xb}.   
 In principle,  no-scale models 
 are necessarily multifield inflation models. 
 This means that apart from the inflaton, there  are additional scalar fields (moduli).

In this paper, we introduce an inflationary model based on the no-scale supergravity \cite{Ellis:2013xoa}. 
Specifically, we consider models with  Starobinsky-like  potential,  derived by no-scale supergravity theories. 
Since,  we need to study the formation of PBHs within these models, we   deform   the 
ordinary $SU(2,1)/SU(2) \times U(1)$ K\"ahler potential,  in order to achieve acceptable  fluctuations to the scalar inflaton  potential, producing  an inflection point. 
For this reason we introduce an exponential term with one extra parameter. We have paid particular attention to satisfy  all the Planck cosmological 
constraints throughout our numerical  analysis. As a result we have found models that satisfying all the phenomenological constraints,  can 
produce up to 20\% of the total DM of the Universe,  due to PBH formation. This value almost saturates the allowed range 
for the PBH abundance, applying  all the relevant   observational data.

The layout of the paper is as follows:  In section~\ref{ballesterpbh} we briefly  review  some basic aspects of supergravity,  relevant to inflation. 
In section~\ref{case1} we modify the K\"ahler  potential and we calculate the effective 
scalar potential by fixing the noncanonical kinetic terms.   We choose  the  inflationary direction and    we verify that  it remains 
stable. Moreover,  we solve the background equation and we justify the insufficiency of the slow-roll approximation. Therefore, we 
describe an algorithm for the numerical solution of M-S equation. Using these solutions we estimate the fractional DM abundance of the PBH as 
a function of its mass 
and we delineate the phenomenologically accepted  regions on this parameter space.
Finally, in section~\ref{conclusion} we give our conclusions and perspectives.

 \section{Supergravity models and inflation}
  \label{ballesterpbh}
The most general $N=1$ supergravity theory is characterized by three
functions. The K\"ahler potential $K$, which is a Hermitian
function of the matter scalar field $\Phi^i$ and describes its geometry,
a holomorphic function of the fields, called superpotential
$W$ and a  holomorphic function $f_{ab}$.

 In the following, we set the reduced Planck mass $M_P= (8 \pi G)^{-1/2}$ to unity. The $N=1$ supergravity action can be written as:
\begin{equation}
\label{01}
\mathcal{S}= \int d^4 x \sqrt{-g} \left( K_{i \bar{j}} \partial_{\mu} \Phi^i \partial ^{\mu} \bar{\Phi} ^{\bar{j}}-V  \right).
\end{equation}
Given  the K\"ahler potential $K$ and the superpotential $W$, one can obtain the  real field metric $K_{i\bar{j}}$ and the scalar potential $V$, following the 
procedure outlined below.

The  general form of  field metric reads as
\begin{equation}
K_{i \bar{j}}(\Phi,\bar{\Phi})= \frac{\partial^2 K}{\partial \Phi^ i \partial \bar{\Phi}^{\bar {j}} }  \,. 
\label{2b}
\end{equation}
Moreover, the scalar potential is given by
\begin{equation}
V= e^{K}\left(	K^{i \bar{j}} \mathcal{D} _i W \mathcal{D}_{\bar {j}} \bar{W}-  3|W|^2\right)+\frac{\tilde g^2}{2}(K^i T^a \Phi_{i})^2\, ,
\label{3}
\end{equation}
where $K^{i \bar{j}} $ is the inverse K\"ahler metric and the covariant derivatives are defined as:
\begin{equation}
\begin{aligned}
& \mathcal{D}_i W  \equiv  \partial_i W + K_i W  \\
& \mathcal{D}^i W   \equiv  \partial^i W - K^i W  \, .
\label{3con}
\end{aligned}
\end{equation}
In addition, we have defined  that $K_i \equiv  \partial K/ \partial \Phi^i$ and, correspondingly,  the complex conjugate $K^i$. 
The last term in the scalar potential (\ref{3}) is just the $D$-term potential, which is set to zero, since the fields $\Phi_i$ are gauge singlets.
From~(\ref{01}) is clear that the kinetic term $\mathcal{L}_{KE}=K_{i \bar{j}} \partial_{\mu} \Phi^i \partial ^{\mu} \bar{\Phi} ^{\bar{j}}$
needs to be  fixed.

The minimal no-scale $SU(1,1) / U(1)$ model is  written in the terms of a single complex scalar field $T$,  with the K\"ahler potential  \cite{Cremmer:1983bf}
\begin{equation}
\label{3i}
K=-3 \ln \left(  T+ \bar{T} \right)\, .
\end{equation} 
In our case,  we consider a no-scale supergravity model with two  chiral superfields $T$, $\varphi$,
that parametrize the noncompact $SU(2,1)/SU(2) \times U(1)$ coset space. 
In this model, the K\"ahler potential can be written as \cite{Ellis:1984bm}
%%%%%%%%%%
\begin{equation}
K=-3 \ln \left( T+ \bar{T} - \frac{\varphi \bar{\varphi}}{3} \right)\, .
\label{1}
\end{equation} 
%%%%%%%%%%%%%%
Then, the corresponding action~(\ref{01}) becomes:
\begin{equation}
\mathcal{S}=\int d ^4x \sqrt{-g} \left[  \frac{1}{2} (\partial_{\mu}\varphi,\partial_{\mu}T) \left( \begin{array}{cc}
 K_{\varphi \bar{\varphi} } & K_{\varphi \bar{T}} \\
K_{T \bar{\varphi} }& K_{T \bar{T} } \\
\end{array}\right)\left( \begin{array}{c}
 \partial^{\mu}\varphi   \\
 \partial^{\mu}T  \\
\end{array} \right)-V(\varphi, T) \right].
\label{8} 
\end{equation}

The simplest globally symmetric model is the Wess-Zumino model, with a single chiral superfield $\varphi$.  
This model is characterized by a mass term $\hat{\mu}$ and a trilinear coupling $\lambda$. Thus,   the superpotential is given by  \cite{Ellis:2013xoa}
 \begin{equation}
W= \frac{\hat \mu}{2} \varphi^2  - \frac{\lambda}{3} \varphi^3.
 \label{1a}
 \end{equation}
 It is possible  to embed this model in the context of  the no-scale $SU(2,1)/SU(2) \times U(1)$ case, by matching the  
  $T$ field   to  the modulus  field and the $\varphi$ to   the inflaton field.  
By doing so, one can derive from (\ref{1}) and (\ref{1a}) a    class of  no-scale models that yield  Starobinsky-like effective potentials. 
This  potential is calculated  along the real inflationary direction defined by 
\beq
T= \bar{T}= \frac{c}{2} \,\, ,  \quad \mathrm{Im}\varphi=0 \, ,
\label{eq:realdir}
 \eeq
 with the choice 
 $\lambda / \mu = 1/3$ and  $\hat{\mu}=\mu \sqrt{c/3}$, where $c$ is a constant.

\begin{figure}[t!]
\centering
\includegraphics[width=130mm]{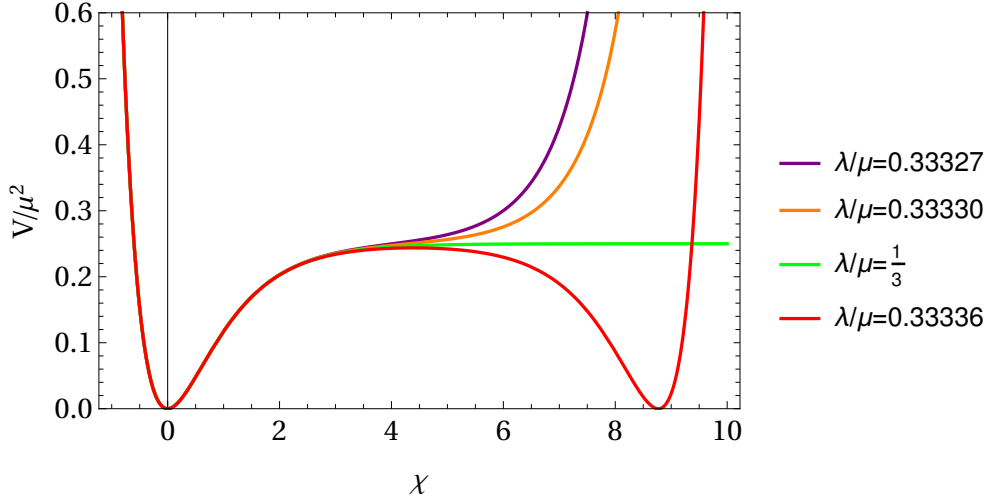} 
\caption{The potential as given by Eqs.~(\ref{1a}) and~(\ref{eq:phichi}) for various values  of the ratio $\lambda / \mu$, as in~\cite{Ellis:2013xoa}.}
\label{f1}
\end{figure}

 In order to have canonical kinetic  terms,  the field $\varphi$ has to be transformed~\cite{Ellis:2013xoa} as 
\begin{equation}
\varphi= \sqrt{3\, c} \,  \tanh \left( \frac{\chi }{\sqrt{3}}  \right) \,  ,
 \label{eq:phichi}
\end{equation}
recovering the potential of the Starobinsky model 
 \begin{equation}
 V (\chi)=\frac{\mu^2}{4} \left(1- e^{-\sqrt{\frac{2}{3}}   \, { \chi} } \right)^2 \, .
 \label{staro}
 \end{equation}
In Fig.~\ref{f1},  we plot the potential  derived  from the superpotential  Eq.~(\ref{1a}) that depends on the ratio $\lambda / \mu$, 
for various values of this ratio around  $1/3$. This central value corresponds to the Starobinsky case. In order to comply with the 
cosmological data~\cite{Akrami:2018odb,Ade:2015lrj,Aghanim:2018eyx} and to explore the dependence on the total number of e-folds, we vary 
the parameter $\mu$  in the range  ($1.8$-- $3.4$)$\times \,10^{-5}$. 

Studying   no-scale models with two chiral superfields   $\varphi$ and  $T$, we notice that  these  fields can interchange roles 
as the inflaton and modulus~\cite{Ellis:2013xoa,Ellis:2013nxa}.
In the  case which $\varphi$ is the modulus field and $T$ is the inflaton, 
the superpotential   reads as \cite{Ellis:2013nxa,Cecotti:1987sa}
\begin{equation}
W=  \mu \, \varphi \, \left(T- \frac{1}{2} \right)\, .
\label{2}
\end{equation} 
   The Starobisky potential is recovered  along the inflationary 
 direction $\varphi=  \bar{\varphi}=\mathrm{Im}T=0$ and $\mathrm{Re}T=\phi$.  In this case too,  in order to have canonical kinetic term, 
 one needs to transform the field $\phi$ to $\chi$ using a relation similar to  (\ref{eq:phichi}).
 Hence, the effective scalar potential is also given by Eq.~(\ref{staro}).

It is essential  one  to verify that the masses of the  inflaton and the modulus  field are not tachyonic.
Thus,  before calculating the evolution of the field,   we  must  check  the stabilization along  the inflationary direction.
If the stabilization is achieved, the modulus  field can be set to be zero and the relevant  term becomes irrelevant to  the dynamical evolution of the inflaton.

\section{Calculating PBH from the modified  K\"ahler potential}
\label{case1}
In this section,  we will study modifications of the K\"ahler potential, that  induce an inflection point 
to  the scalar potential, and consequently causes  peaks in the CMB power spectrum.
For this reason, we use as basis   the  Wess-Zumino potential (\ref{1a}),  modifying  the K\"ahler potential, by introducing  an exponential term as 
\begin{equation}
K= -3 \ln  \left[ T +\bar{T} -\frac{\varphi \bar{\varphi}}{3}+a \, e^{-b(\varphi+ \bar{\varphi})^2}(\varphi+\bar{\varphi} )^4 \right] \, ,
\label{a1}
\end{equation}
 \noindent
where $a$ and $b$ are real numbers. 
Obviously, in the limit  $a=0$, we retrieve the result that corresponds to the  Starobinsky potential, as calculated  in the previous section. 
Moreover, expanding the exponential, one obtains a polynomial modification of the K\"ahler potential, 
as it has been used in the literature~\cite{Garcia-Bellido:2017mdw,Hertzberg:2017dkh,Gong:2017qlj}. 
The particular  exponential form has the advantage that practically   introduces   just one extra parameter,  $b$. 
 In our analysis the parameter $a$ gets just 
two values: $a=0 $  to switch off the effect of the modified term  and  $a=-4 $ when the extra term is used.

The real part of the field $\varphi$ plays the role of the inflaton. 
In order to verify   the  stability of the potential, along the real direction  in Eq.~ (\ref{eq:realdir}),
 we   calculate  the squared mass  matrix and we check 
  that no  tachyonic instability is present, that is  $m^2_{\mathrm{Re}T},m^2_{\mathrm{Im}T},m^2_{\mathrm{Im}\varphi} \ge	0$. %,  along this  inflationary direction. 

In detail,  the general form of mass matrix is 
\begin{equation}
m^2_{s}=\begin{pmatrix} 
(K^{-1})^i_k \mathcal{D}^k \partial_j V & (K^{-1})^i_k \mathcal{D}^k \partial^j V \\
 (K^{-1})^k_i \mathcal{D}_k \partial_j V  & (K^{-1})^k_i \mathcal{D}_k \partial^j V
\end{pmatrix} \,, 
\end{equation} 
\noindent
where $(K^{-1})^i_j$ is the inverse metric of $K^j_i=\partial ^2 K / \partial \Phi^i \partial \Phi_j$ and the K\"ahler covariant derivative is given in (\ref{3con}).
Specifically, in the  case  of the  two chiral fields the mass matrix takes the form     
\begin{equation}
m^2_{s}=\begin{pmatrix} 
(K^{-1})^\varphi_{k}\mathcal{D}^{k} \partial_ {\varphi} V & (K^{-1})^\varphi_{k}\mathcal{D}^{k} \partial^{\bar{T}} V  \\
(K^{-1})^{k}_{{T}}\mathcal{D}_{k} \partial_{\bar{\varphi}} V   & (K^{-1})^k_{\bar{T}}\mathcal{D}_{k} \partial^{\bar T} V \end{pmatrix}.
\label{massmatrix1}
\end{equation}

Following~\cite{Ellis:2019bmm,Ellis:2018xdr}, we have computed analytically and numerically the masses of the fields $\varphi$ and $T$ and we have verified that along the real direction, 
  $T= \bar{T}$ and $\varphi=\bar{\varphi}$,  the eigenstates of the matrix~(\ref{massmatrix1}) are positive.  
  Unfortunately, the corresponding equations are too long  to be displayed here.
Repeating the same calculation in the  imaginary direction, we have checked the positivity of the mass eigenstates,
using   $\vev{\mathrm{Re} T}= \frac{c}{2}$ and $\vev{ \varphi} =0$.

Having verified   the stability along the inflationary direction, using  Eqs.~(\ref{2b}),(\ref{3}),   the scalar effective potential can be calculated. As a first step,
we find   the field transformation,   that puts the kinetic term in canonical form. 
Moreover, defining  $\mathrm{Re}\,  \varphi \equiv \, \phi$, the relevant term in Eq.~(\ref{8}) 
is the    $K _{\varphi \bar{\varphi}}$, which  along the direction (\ref{eq:realdir}), apparently equals to $K_{\phi \phi}$. Thus,   one gets 
\beq
\frac{1}{2} \, \partial_{\mu}\chi \,  \partial^{\mu}\chi= K_{\phi \phi}  \, \partial_{\mu}\phi  \, \partial^{\mu}\phi \,
\eeq
or equivalently 
\beq
\frac{d \chi}{d \phi}= \sqrt{2 K_{\phi \phi }}\, .
\label{9case3}
\eeq
By integrating the latter,  we obtain the generalization of Eq.~(\ref{eq:phichi}), using appropriate boundary conditions. 
These conditions  are fixed from  the requirement   to  retrieve   the  Strarobinsky case, in the limit $a = 0$.

Afterwards, we compute the scalar potential along the direction (\ref{eq:realdir}), using Eq.~(\ref{3}) and the modified K\"ahler potential from (\ref{a1}), as
\begin{equation}
V(\phi)= \frac{3 e ^ {3 b \phi^2} \phi^2(c \mu^2 -2\sqrt{3c} \lambda \,  \mu \,  \phi+3 \lambda^2 \,  \phi^2)}
     {\left[  -3 a \phi^4 +e ^{b\phi^2}(-3c+\phi^2) \right]^2\, \left[ e^{b\phi^2} -6\, a \, \phi^2(6+b \, \phi^2(-9+2 b \, \phi^2)) \right]} \,.
\label{eq:vofphi}
\end{equation}
Finally, using the  generalized relation   $\phi(\chi)$, obtained by  Eq.~(\ref{9case3}), the potential above can be expressed as $V(\chi)$. The precise   form of the 
 $V(\chi)$ is obtained only numerically, due to its complexity and this numerical relation is used thereafter.

\begin{table}[t!]
\centering
 \begin{tabular}{||c c c ||} 
  \hline
  &$\lambda / \mu$  & $b$  \\ [0.5ex] 
 \hline\hline
 1.&$0.33327$ &$87.379427$\\ 
\hline
  2.&$0.33330$  & $87.390563$\\ 
 \hline
  3.&$1/3$  &$87.402941$\\ 
 \hline
 \end{tabular}
 \caption{ The values of the parameters  $\lambda/\mu$ and  $b$,
 for $a=-4$ and   $2\vev{\mathrm{Re}T}= c= 0.065 $.}
 \label{tabu4}
\end{table}
\noindent

 \begin{figure}[t!]
\centering
\includegraphics[width=115mm]{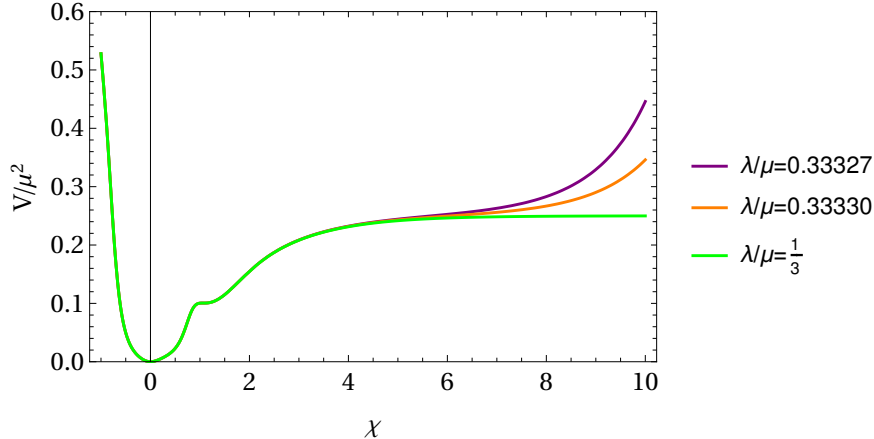} 
\caption{The potential given in Eq.~(\ref{eq:vofphi}) as a function of $\chi$,  
for various values of the ratio $\lambda  / \mu$ as in  Table~\ref{tabu4}.}
\label{f4}
\end{figure}

In Fig.~\ref{f4} we plot the potential $V(\chi)/\mu^2$, as a function of the field $\chi$,
 using the values of the parameters $\lambda/\mu$ and  $b$,  as  in  Table~\ref{tabu4}. 
  The parameter   $\mu  $ is fixed in order to satisfy the Planck constraint for power spectrum, 
which is approximately  $P_\mathcal{R}=2.1 \times 10^{-9}$, at a pivot scale of $k_*=0.05 \, \mathrm{Mpc}^{-1}$. As we will discuss below, 
varying  the parameter $\lambda$ affects mainly the spectral index $n_s $, but also the   tensor-to-scalar   ratio $r$ of the power spectra. 
After fixing   $\lambda$ and $\mu$, the  values for  $b$ in  Table~\ref{tabu4}, are chosen  in order the PBH abundance to saturate the cosmological bounds. 
which as  we will see,  constrain significantly the parameter space of the PBH.
 The prediction of the model is not very sensitive on the $a$, and thus is chosen to be  $a=-4$.
  Finally, in the context of our model,
   the parameter $c$ affects mainly the total number of e-folds. 
   To get agreement with the Planck 2018 data we choose  $c=0.065$\footnote{In the original  model based on the K\"ahler potential as in Eq.~(\ref{1}),
   the dependence on the parameter $c$ drops out~\cite{Ellis:2013xoa}.  In particular,  this results from  the transformation in  Eq.~(\ref{eq:phichi}) and the 
   redefinition $\hat{\mu} \to \mu \sqrt{c/3}$. In the context of the modified K\"ahler potential ~(\ref{a1}) there is indeed  a remaining $c$-dependence, that is fixed by the Planck data.}. 

One can notice,  that the potential  has the required  features  that ensure that sizable abundance of  PBH is created.
 Specifically, the potential around the inflection point $\chi \sim 1 $,   satisfies the relations 
 \begin{equation*}
\frac{dV(\chi_{i})}{d\chi_{i}} \simeq 0,  \quad \frac{d^2V(\chi_{i})}{d\chi_{i}^2}=0 \, .
 \end{equation*}
 Around the inflection point,  the inflaton slows down, generating  a large amplification in the power spectrum. 
In addition, it has a minimum with $V(\chi_0)=0$, at $\chi_0=0 $,  to achieve the 
reheating,    after inflation ends.      
 
 \begin{figure}[h!]
\centering
\includegraphics[width=95mm]{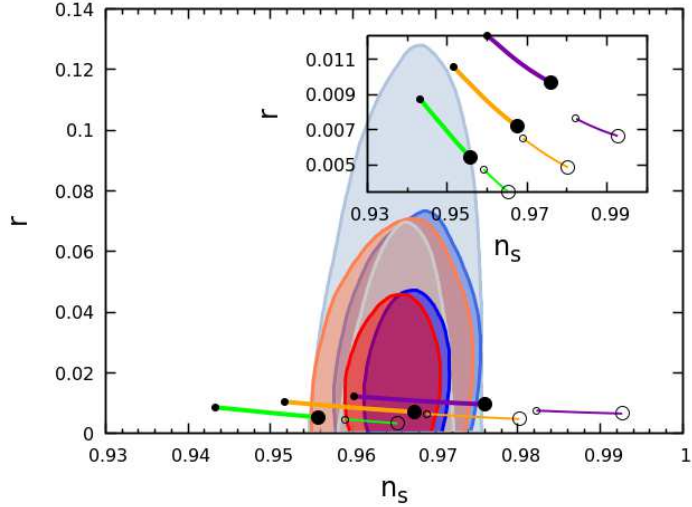} 
\caption{The predictions of our model  for the tilt $n_s $  and  the tensor-to-scalar ratio $r$. The shaded regions are taken  from 
Planck 2018 and other data \cite{Akrami:2018odb}. For the details see the main text.
% from Planck 2018 \cite{Akrami:2018odb}. For the contour details are discussed in  \cite{Akrami:2018odb}.
%alone $TT,TE,EE+lowE+lensing$ (gray) and in combination with $TT,TE,EE+lowE+lensing+BK15$ (red) and $TT,TE,EE+lowE+lensing+BA0$ (blue) with marginalized joint $68\%$ and $95\%$ CL.
 }
%\textcolor{red}{\bf move the most of the caption text in the main text.}}
\label{f6}
\end{figure} 
 
 In Fig.~\ref{f6}  we plot  the predictions for the tilt $n_s$ in the spectral index of scalar perturbations and for the tensor-to-scalar ratio $r$,
 of the original Wess-Zumino model (thin line segments with empty dots) and the model with modified K\"ahler potential (thick line segments with filled dots), compared against 
   the recent  data of Planck 2018, that prefer  the central shaded regions in the plot.
   The meaning of the colors of these regions are explained in the Planck collaboration analysis~\cite{Akrami:2018odb}. 
   %For comparison reasons,    we plot  the results for  the Wess-Zumino model by~\cite{Ellis:2013xoa} (thin lines). 
  %Furthermore, we depict the results of this analysis, assuming the modified K\"ahler potential, which lead to the formation of PBH (thick lines). 
  Green colored lines  correspond to the case $\lambda / \mu= 1/3$, the orange to $0.33330$ and the purple to $0.33327$. 
  The evolution of the field is fixed by requiring  50 (small dots), or 60 (big dots) e-folds at the end of the line segments. 
We  notice,  that introducing the  modified potential in Eq.~(\ref{a1}), the cosmological  predictions are affected considerably. 
  Therefore, some values  of the ratio $\lambda / \mu$,  which were  originally excluded,  become acceptable in the modified case.
% One can derive similar results by using the rest set of parameters in Table~\ref{tabu4}. 

\subsection{Applying  the slow-roll approximation}
  \label{rpbh}
  
 The evolution of the inflaton  field $\chi$ in a Friedmann–Robertson–Walker (FRW)
 homogeneous background, which we take to be spatially flat, is driven by the system of the Friedmann equation and the inflaton field equation:
 \begin{equation}
\label{20a}
\begin{split}
H^2= \frac{1}{3} \left(\frac{1}{2} \dot{\chi^2}+V(\chi)\right)\\
\ddot{\chi}+3H\dot{\chi}+V'(\chi)=0 \,, 
\end{split}
\end{equation} 
\noindent
where dots represent derivatives with respect to cosmic time and primes the derivatives with respect to the field $\chi$.
We can rewrite the system above in terms of number of e- folds elapsed from initial cosmic time $t_i$ described by the integral:
\begin{center}
$N(t)=\int ^t _{t_i} H(t') dt'$.
\end{center} 
\noindent
So,  the background equation or the equation of the inflaton field take the form
\begin{equation}
\label{21}
\frac{d^2 \chi}{d N^2} +3 \frac{d \chi}{dN}- \frac{1}{2} \left(\frac{d \chi}{d N}\right)^3 +\left[ 3- \frac{1}{2} \left(\frac{d \chi }{d 	N}\right)^2\right] \frac{d \ln V(\chi)}{d \chi}=0.
\end{equation}

We solve  numerically the Eq.~(\ref{21}), using as initial conditions those that, in the slow-roll approximation   are compatible with the cosmologically 
acceptable values for $n_s$ and $r$~\cite{Ade:2015lrj,Akrami:2018odb,Aghanim:2018eyx}.
Specifically, by  the Planck 2018  data~\cite{Akrami:2018odb}  on inflationary parameters, at the pivot scale $k_*=0.05\,  \mathrm{Mpc}^{-1}$, we get
\begin{equation}
\begin{array}{cc}
n_s=0.9625 \pm 0.0048 \\
r<0.044.
\end{array}
\label{Planck2015}
\end{equation}
  We evaluate the spectral index $n_s$ 
 and the tensor-to-scalar   ratio $r$, at leading order in the slow-roll expansion by
\begin{equation}
\label{21a}
n_s \simeq 1+2 \eta_V -6\varepsilon_V, \quad r \simeq 16 \varepsilon_V \, , 
\end{equation} 
where the relevant slow-roll parameters are defined as
\beq
\varepsilon_V= \frac{1}{2}\left(\frac{V'(\chi)}{V(\chi)} \right)^2 , \quad \eta_V=\frac{V''(\chi)}{V(\chi)}.
\eeq
Using the numerical relation between  $\phi$ and  $\chi$, based on Eq.~(\ref{9case3}), 
 the  initial condition for the field $\phi$  
  can be transformed to the initial conditions for the  $\chi$.  As for the initial condition for the  derivative of $\chi$, we use 
    the slow-roll attractor relation 
    \beq
    \label{eq:attract}
     \frac{d \chi}{dN} \approx - \left| {\frac{dV}{V d\chi}}  \right| \, .
     \eeq
Consequently,  the numerical solution  for the  slow-roll parameters reads as 
\begin{equation}
 \label{22}
\varepsilon_H=\frac{1}{2} \left( \frac{d \chi}{dN} \right)^2, \quad \eta_H= \varepsilon_H- \frac{1}{2} \frac{d \ln\varepsilon_H}{dN}.
\end{equation}
Using this equation for $\varepsilon_H$,  the Hubble function squared reads from Eq.~(\ref{20a}) as
\begin{equation}
 \label{23}
H^2=\frac{V(\chi)}{3- \varepsilon_H}.
\end{equation}
Given  these expressions, we evaluate the power spectrum within the slow-roll approximation, as:
\begin{equation}
 \label{24}
P_\mathcal{R} \simeq \frac{1}{8 \pi^2} \frac{H^2}{\varepsilon_H}.
\end{equation}
Notice  that for the numerical solution of the background Eq.~(\ref{21}), one must use   the Eq.~(\ref{9case3}) and Eq.~(\ref{eq:vofphi}). 
As usual, the condition $\varepsilon_H \approx 1$ marks the end of inflation
and  the numerical calculation ends at this point. 
We constrain the number of e-folds $N$, that is the number of e-folds elapsed  between the time that today's largest observable 
scales exit the Hubble horizon  and the time at which inflation ends, to be   $45- 55$.

\begin{table}[t!]
\centering
 \begin{tabular}{||c c c | c c||} 
 \hline
  &$\phi_0$  & $\lambda/ \mu$   & $n_s$ & $r$ \\ [0.5ex] 
 \hline\hline
 1.& $0.4258$& $0.33327$ &$0.961234$& $0.0121106$ \\ 
\hline
  2.& $0.4272$ & $0.33327$ & $0.967463$ & $0.0109205$ \\ 
   \hline
  3.&$0.4266$ & $0.33330$ & $0.958265$ & $0.00900217$\\ 
 \hline
  4.&$0.4258$ & $1/3$ &$0.948072$  & $0.00740699$\\ 
\hline
 \end{tabular}
 \caption{ The initial conditions for the field $\phi$,  $n_s$ and $r$ that correspond  to sets in  Table~\ref{tabu4}. (The  first two sets 
 correspond to the first set of Table~\ref{tabu4}, with different $\phi_0$.)    }
 \label{tabu2}
\end{table}
\noindent
 
\begin{figure}[t!]
\centering
\includegraphics[width=80mm,height= 64mm]{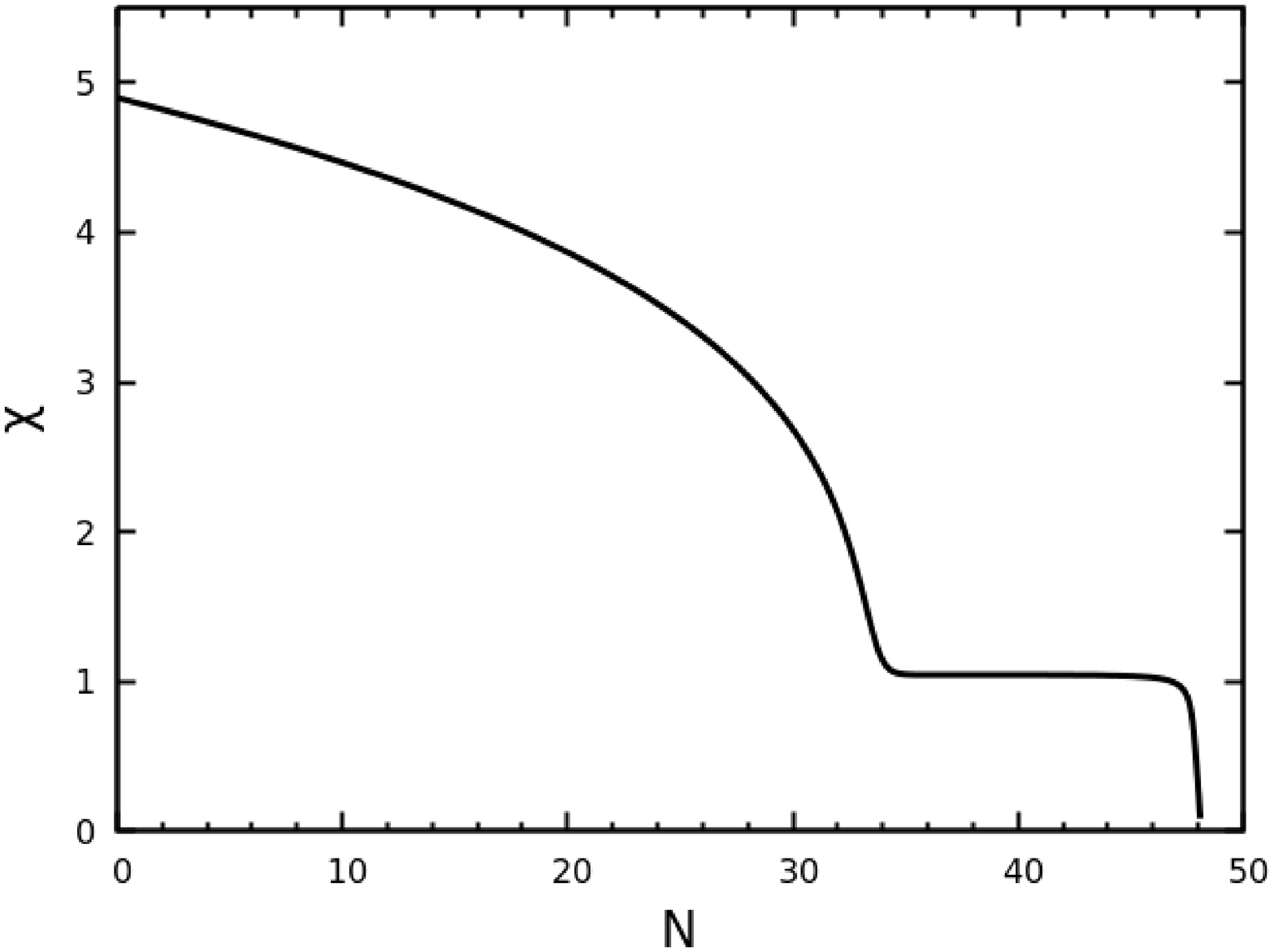} \, \includegraphics[width=80mm]{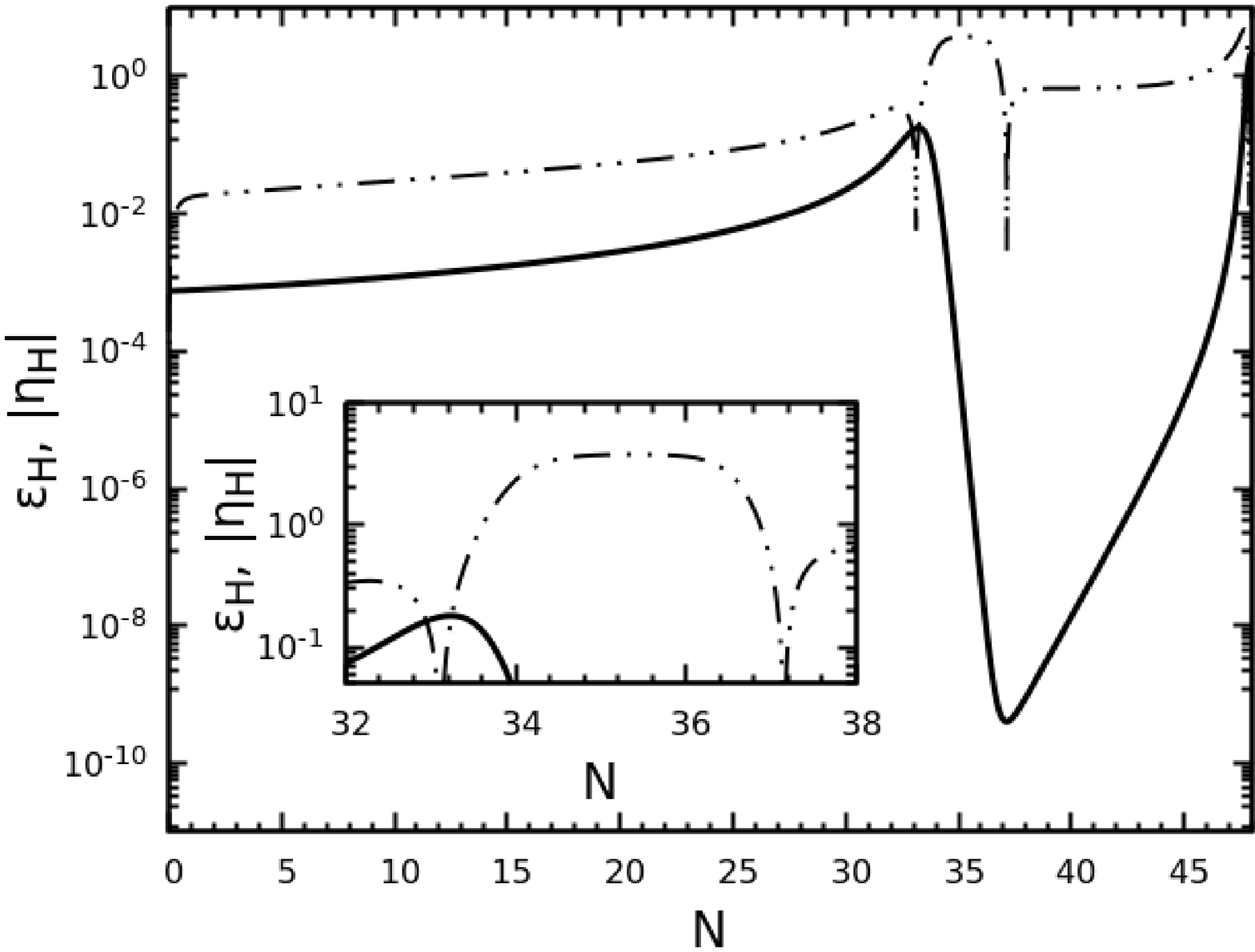} 

\caption{Left panel: The evolution of the inflaton field $\chi$, in Planck units, as a function of the numbers of  e-folds.
Right panel: The slow-roll parameters using the first set of parameters in Table~\ref{tabu4}. Solid line corresponds to $\varepsilon_H$ and dashed to $|\eta_H|$.}
%\textcolor{red}{\bf which field }
\label{f5}
\end{figure}

% We have modified the K\"ahler potential in order to achieve significant peaks in the power spectrum. 
 In our numerical analysis,   we use the sets of parameters given in Table~\ref{tabu4}, as discussed in the beginning of this section. 
 For the initial condition of the field $\phi$, $\phi_0$, we use the numbers in the first column in  
 Table~\ref{tabu2}. Please note that, the first two lines  in Table~\ref{tabu2}, correspond to the first line in Table~\ref{tabu4}.
The last two columns in Table~\ref{tabu2} are the outcome of the calculation,   the predicted  values for the observables $n_s$ and $r$.
 The initial conditions $\phi_0$ and $N_*=0$  are set to the point  that CMB scales cross the horizon. 
 In addition, 
 this point  corresponds to the asymptotic plateau of the potential $V(\chi)$ in Fig.~\ref{f4}. 
%The initial value of the field is taken in order to be in accordance with the Planck 2018 data \cite{Akrami:2018odb}. 
 At the end of this procedure, we  calculate the evolution of the field $\chi$ and  the slow-roll parameters $\varepsilon_H$, $\eta_H$ in terms of $N$,
  and show our results in  Figs.~\ref{f5} left and right panel, respectively.

As it can be seen in Fig.~\ref{f5} right,  the value of parameter $\varepsilon_H$ remains always 
below 1 until the end of inflation.  We  further notice that the inflaton reaches the region of reheating
 at the global minimum of potential in Fig.~\ref{f4}, that corresponds to $N\simeq 50$ in Fig.~\ref{f5} left, as it was expected.

 Using the slow-roll approximation for calculating the power spectrum as in Eq.~(\ref{24}), we can get sizable peaks. 
  However, paying attention to  the details of the slow-roll approximation, especially to the values of the parameters $\varepsilon_H$ and $\eta_H$ 
 in Fig.~\ref{f5} right, we remark that they  get values of $\mathcal{O}(10^{-1})$--$\mathcal{O}(1)$, that clearly violate this approximation. 
Therefore,  it is crucial  to solve the precise  M-S equation and then we can proceed to the evaluation of fractional abundance of PBH.

\subsection{Solving the Mukhanov-Sasaki equation}
  \label{vabrfd}
 As it has been explained in the previous section the slow-roll 
approximation fails to reproduce the correct power spectrum and hence the correct mass of PBH as well as, the fractional abundances. The fact that the values of slow-roll parameters $\varepsilon_H$ and $\eta_H$ are close to 1 and over 3 
respectively, leads us to search for a more accurate method. When the potential has a sharp feature such as an inflection point, it is crucial to 
 evolve the full mode equation numerically, without any approximation \cite{Adams:2001vc}. Hence, we need to have an precise  solution of the power spectrum, versus the
 comoving wave-number $k$ in order to produce the abundance of PBH.  This solution can be found by the so-called 
 M-S equation\cite{Mukhanov:1988jd,Sasaki:1986hm} 
  which is given by the following expression:
\begin{equation}
 \label{25}
\frac{d^2u_k}{dN^2}+(1-\varepsilon_H) \frac{du_k}{dN}+\left[  \frac{k^2}{\mathcal{H}^2}+(1+\varepsilon_H-\eta_H)(\eta_H-2)- \frac{d(\varepsilon_H- \eta_H)}{dN} \right]u_k=0
\end{equation}
and
\begin{equation}
 \label{26}
u= z R , \quad z = \frac{a}{\mathcal{H}} \frac{d \phi}{d \tau},
\end{equation}
\noindent
where $R$ is the comoving curvature perturbation and $a$ is the scale factor. 
We denote by $\tau$ the conformal time and by $\mathcal{H}=a H$ the conformal Hubble parameter.
 Instead of working with complex coefficients, it is convenient to solve the M-S 
equation twice: one  for the real and one for the imaginary part for each mode $u_k$. The corresponding  initial conditions are \cite{Adams:2001vc}:
\begin{equation}
\begin{aligned}
& \mathrm{Re} \left(u_k \right)=\frac{1}{\sqrt{2 k}} \, ,\quad \mathrm{ Im} \left( u_k \right)=0 \\
& \mathrm{Re} \left( \frac{du_k}{dN_i} \right)=0  , \quad \mathrm{Im} \left( \frac{du_k}{dN_i}  \right)=-\frac{\sqrt{k}}{\sqrt{2}k_i}
\end{aligned}
\label{26b}
\end{equation}
where $k_i$ is chosen a thousand times smaller than the wave-number of interest.
To evaluate the power spectrum we repeat the integration over  many values of $k$. The numerical precise  value of spectrum (solving the M-S equation) is given by:
\noindent
\begin{equation}
\label{27}
P_\mathcal{R}= \frac{k^3}{2 \pi^2}{ \Big| {\frac{u_k}{z}\Big|^2 }} _{k \ll \mathcal{H}}.
\end{equation}
%We note that the background equations contribute to this in $\varepsilon_H$ and $\eta_H$

\begin{figure}[t!]
\centering
\includegraphics[width=100mm]{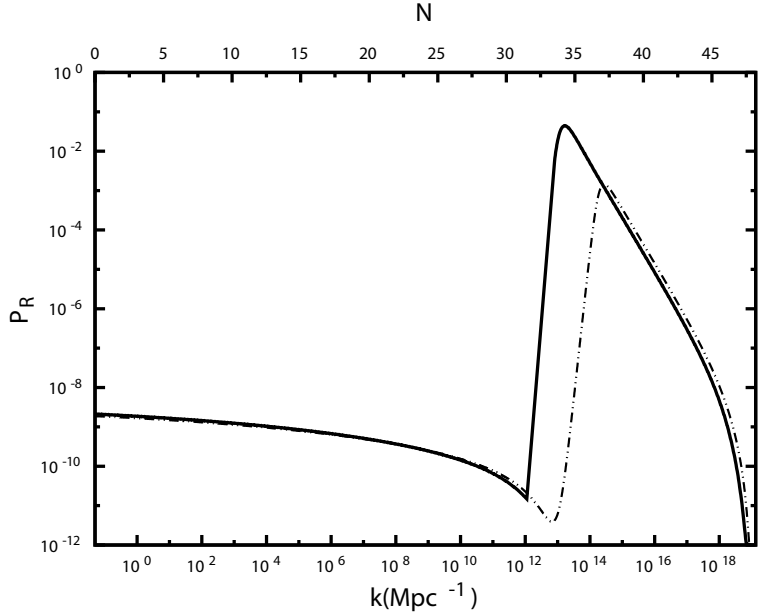} 
\caption{The CMB power spectrum using the slow-roll approximation (dashed line) and  the M-S formalism  (solid line) for the first set of parameters of Table~\ref{tabu4}.}
\label{f8}
\end{figure}

The numerical strategy for solving the M-S equation, based on refs. \cite{Ballesteros:2017fsr,Mahbub:2019uhl}, is summarized below:

\begin{itemize}
\item{The background Eq.~(\ref{21}) is solved numerically using the initial conditions for the field and its first derivative. The numerical solution stops when the condition $\varepsilon_H=1$  is satisfied, denoting the end of inflation. The total number of e-folds is defined between the times where the $k$-modes exit and enter the Hubble horizon. The transformation of the field needs to be taken into account too.}
\item {In order to solve the  Eq.~(\ref{25})   the solution of  the background equation for  $\chi$  is required, as well as the slow-roll parameters from the previous steps. 
The second and third derivatives of $\chi$ in the last term of (\ref{25}) are also be evaluated using Eq.~(\ref{21}) and its first derivative with respect to $N$.}
\item{One can now solve the M-S equation. For each mode of interest $k$,  the  Eq.~(\ref{25}) is solved  twice with the initial conditions given by (\ref{26b}), until the solution is approximately constant ($\frac{u_k}{z} \approx const$). We choose the values of initial $N_i$  to be  $ N=N_i-N_* $ and the connection between the number of e-folds and the comoving wave-number is given by: 
\begin{equation}
 k=k_* \frac{H(N_i)}{H(N_*)} e^{N_i-N_*} \, .
\end{equation}
%as it has been proposed in ref. \cite{Ballesteros:2017fsr}
The initial value of $k_*$    is $k_*=0.05\, \mathrm{Mpc}^{-1}$ and we assume that $N_*=0$, as the CMB scales exit the Hubble horizon.}
% (We can choose the values of $N_e$ arbitrarily by associating one $k$ with one $N$.)} 
\item{Eventually,  the  $P_\mathcal{R}$ is evaluated  precisely using  Eq.~(\ref{27})
for each $k$-mode of interest, which is related to $N$ as it is explained in
the previous step.
As for the normalization of the  power spectrum
we use that it is approximately $2.1\times10^{-9}$ \cite{Aghanim:2018eyx}
at $k_*=0.05 \, \mathrm{Mpc}^{-1}$. }
\end{itemize}

With this algorithm we are able  to reproduce previous works,
 such as those of refs.\cite{Ballesteros:2017fsr,Ozsoy:2018flq,Gao:2018pvq,Cicoli:2018asa}. This numerical method is applied
 to our case, where the K\"ahler potential is modified. The power spectrum is evaluated
 using  Eqs.~(\ref{25}) and (\ref{27}) and depicted in Fig.~\ref{f8} for the first set of 
 parameters shown in Table~\ref{tabu4} taking into consideration that the initial condition for the background equation is given by the first set of  Table~\ref{tabu2}. 
 The solid line corresponds to the M-S power spectrum and the dashed line to the slow-roll approximation as in Eq.~(\ref{24}).
  As one can notice in   Fig.~\ref{f8},
  despite the fact that  peaks can be  produced  within the 
 slow-roll approximation, this approximation fails to reproduce either the  peak's   height or its position.
 The numerical precise  result of power spectrum ensures that the value of peak's height is
   larger  than $10^{-2}$ and hence a significant fractional abundance of PBH can be achieved, as it is shown in the next section.

We notice that   employing improvements of the slow-roll approximation like the optimized slow-roll approximation~\cite{Motohashi:2017kbs},   
the size of power spectrum   peak,  approaches indeed  this of the M-S numerical solution. 
On the other hand,  although  using  either the slow-roll or its improvement, the peak's position is not affected, this is 
quite different from this of the numerical solution. 
  As it will be discussed below, since   the position of the peak  is crucial for the precise calculation of fractional PBH abundance,
in the following  we will use the M-S numerical solution, as it is suggested in~\cite{Germani:2017bcs}.

\subsection{The calculation of the  PBH abundance}
  \label{abundance}

Using the precise calculation of the 
 power spectrum  via the M-S equation, as described in the previous section, we can evaluate  the fractional abundance of PBH that can be interpreted as  DM. 
 For this reason, we will employ the  Press-Schechter model,    that is  used in   the gravitation collapse \cite{Press:1973iz}. This model is summarized below.
%\begin{itemize}
%\item

First, we need to compute the coarse-grained mass variance, which is defined in the radiation-dominated era  as:
\begin{equation}
\label{40}
\sigma^2 \left( M(k)  \right)= \frac{16}{81}  \int \frac{dk' }{k'} \left(\frac{k'}{k}\right)^4 P_\mathcal{R}(k') W^2\left(\frac{k'}{k}\right),
\end{equation}
where $W(x)=e^{ -x^2/2}$ is the Gaussian distribution. 
Knowing  $\sigma(M(k))$ we evaluate the mass fraction of PBH at formation,  denoted by $\beta(M)$:
\begin{equation}
\label{42}
\beta(M)= \frac{1}{\sqrt{2 \pi \sigma ^2 (M)}} \int^{\infty}_{\delta_c} d\delta \,  \exp \left(  -\frac{\delta ^2}{2 \sigma^2(M) } \right).
\end{equation}
%If $\sigma$ is above a certain threshold the probability of forming PBH increases. 
The value of $\delta_c$, which denotes the critical value for collapse to produce a PBH,  plays a crucial role in this procedure. 
%then the probability of collapsing to form a PBH can be large. ,where $\delta_c$ denotes the critical value for collapse to produce a PBH, 
The integral in Eq.~\ref{42} is evaluated using the incomplete gamma function 
\begin{equation}
\beta(M)=\frac{\Gamma(\frac{1}{2}, \frac{\delta_c^2}{2\, \sigma^2(M)})}{2\sqrt{\pi}} \, .
\label{42b}
\end{equation}
%\item
As the next step we compute the mass as a function of $k$ \cite{Ballesteros:2017fsr}:
\begin{equation}
\label{43}
M(k)=10^{18} \left( \frac{\gamma}{0.2} \right) \left(\frac{g_*(T_f)}{106.75}\right)^{-1/6} \left(\frac{k}{7 \times 10^{13} \, \mathrm{Mpc}^{-1}  }\right)^{-2}  \mathrm{ in~grams} \, .
\end{equation}
%\textcolor{red}{\bf Define $g$ is the eff dof? }\textsl{g}
This expression runs over  all the $k$-modes. With $\gamma$ we denote a factor which depends on gravitation collapse
and we choose $\gamma =0.2$~\cite{Carr:1975qj}.
 $T_f$   denotes  the temperature  of PBH formation.    $g_*(T_f)$  are the effective degrees of freedom during this formation and counting only the 
 SM particles we set $g_*(T_f)=106.75$.

 %, which is equal to entropy density, meaning $g(T_f)=106.75$ \cite{Ballesteros:2017fsr}.
%\item
Given the mass fraction $\beta$ and the mass $M(k)$ we can evaluate the abundance $\Omega_ {PBH} / \Omega_ {DM}$ as a function of mass
%The equation is given by the following expression:
\begin{equation}
\label{44}
\frac{\Omega_ {PBH}}{\Omega_ {DM}}(M)= 
\frac{\beta(M)}{8 \times 10^{-16}} \left(\frac{\gamma}{0.2}\right)^{3/2} \left(\frac{g_*(T_f)}{106.75}\right)^{-1/4}\left(\frac{M \,  }{10^{-18 }\; \mathrm{ grams} }\right)^{-1/2}.
\end{equation}
Hence, we plot $\frac{\Omega_ {PBH}}{\Omega_ {DM}}(M)$ versus $M(k)$.
%\item
Finally, we integrate the expression in Eq.~(\ref{44}) as
\beq
\Omega_ {PBH} = \int \frac{d  M}{M} \, \Omega_ {PBH} (M) \, ,
\eeq
 in order to find the present abundance  and  the results are in  Table~\ref{tabu42}. 
%\end{itemize}

\begin{table}[t!]
\centering
 \begin{tabular}{||c c c c ||} 
  \hline
  & $P^{peak}_{\mathcal{R}}$& $M^{peak}_{PBH}/ M_\odot$  &$\Omega_{PBH} / \Omega_{DM} $ \\ [0.5ex] 
 \hline\hline
1.& $4.472\times 10^{-2}$&$5.544 \times 10^{-14}$ & $0.165$ \\
2.& $3.968\times 10^{-2}$ & $1.171 \times 10^{-16}$&  $0.095$   \\
3.& $3.988\times 10^{-2}$ & $7.399 \times 10^{-17}$&  $0.121$    \\%Pote
4.& $3.998\times 10^{-2}$ & $8.787 \times 10^{-17}$&  $0.121$     \\
 \hline
 \end{tabular}
  \caption{ The values of the peak of power spectrum using  $\delta_c=0.45 $ and their fractional abundance,  which correspond  to the parameter sets
   in Table~ \ref{tabu2}.}
 \label{tabu42}
\end{table}

\begin{figure}[t!]
\centering
\includegraphics[width=110mm]{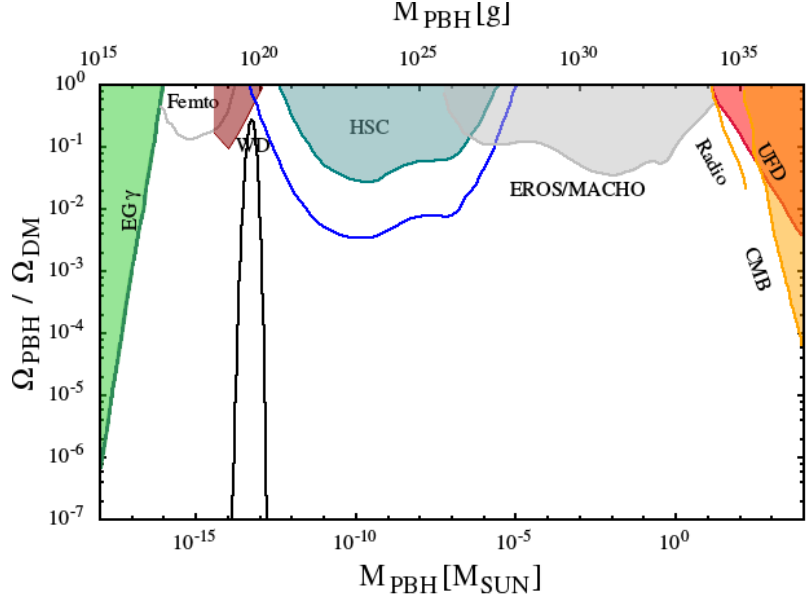} 

\caption{The fractional abundance of PBH for the first set of parameters in
Table~\ref{tabu2} (black line). Details on the various
excluded regions  due to  observation data given in
\cite{Carr:2009jm,Barnacka:2012bm,Graham:2015apa,Capela:2013yf,Niikura:2017zjd,Tisserand:2006zx,Monroy-Rodriguez:2014ula,Brandt:2016aco,Koushiappas:2017chw,Ali-Haimoud:2016mbv,Poulin:2017bwe,Gaggero:2016dpq,Inoue:2017csr}. }
\label{f9}
\end{figure}

\begin{figure}[t!]
\centering
\includegraphics[width=110mm]{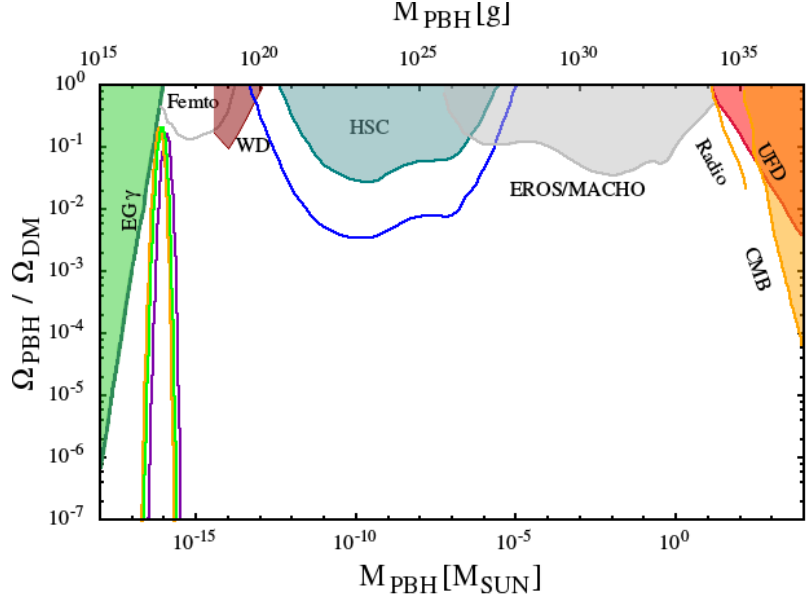} 

\caption{The fractional abundance of PBH, such as Fig.~\ref{f9}, for the last three sets of parameters in Table~\ref{tabu2}.}
\label{f10}
\end{figure}

One  should notice  that the  calculated  PBH abundance   is  sensitive to the value of $\delta_c$.
This value  depends on the profile of the collapsing overdensities.  
Recent studies, assuming   radiation domination, 
 suggest that  $\delta_c \approx 0.4-0.5$ \cite{Harada:2013epa,Musco:2012au,Chisholm:2006qc,Musco:2008hv,Musco:2004ak,Escriva:2019phb,Escriva:2020tak,Germani:2018jgr}. 
 The same result is supported, by   analytical calculations~\cite{Escriva:2019phb,Escriva:2020tak} .
  Furthermore,  one can notice by Eqs.~(\ref{40}) and(\ref{42b}) that the 
PBH abundance depends also  on the value of the 
  the power spectrum peak, since $\sigma$ is in the denominator in the exponential. 
  This is an additional  justification for employing the precise numerical solution
  of the  M-S equation, instead of the   slow-roll approximation.

Using  this method, we are able to  produce a significant abundance of PBH, modifying accordingly   the K\"ahler potential, through the $b$ parameter. 
We summarize our results in Table~\ref{tabu42}, where we have used $\delta_c = 0.45$. 
The sets of parameters in this table correspond to those in Table~\ref{tabu2}. 
We must stress that the amount of the fine-tuning in   the parameter $b$, is related to the central  value for the $\delta_c$ we have used.
For example, using $\delta_c = 0.5$, the values of $b$ as appear in Table~\ref{tabu4} are less tuned in the last two digits.   
This means that allowing a slight  variation on $\delta_c$, we can somewhat reduce the fine-tuning on  $b$.

We plot the fractional abundance for the first set of parameters of Table~\ref{tabu4} (Fig.~\ref{f9}). 
The observational data depicted in Figs.~\ref{f9} and \ref{f10} are adapted by\cite{Ballesteros:2017fsr} with the bounds by refs\cite{Carr:2009jm,Barnacka:2012bm,Graham:2015apa,Capela:2013yf,Niikura:2017zjd,Tisserand:2006zx,Monroy-Rodriguez:2014ula,Brandt:2016aco,Koushiappas:2017chw,Ali-Haimoud:2016mbv,Poulin:2017bwe,Gaggero:2016dpq,Inoue:2017csr}.
 Specifically, these bounds are from extragalactic gamma ray from PBH evaporation ($EG\gamma$) \cite{Carr:2009jm}, femtolensing of gamma ray burst (Femto) \cite{Barnacka:2012bm}, white dwarfs explosion (WD) \cite{Graham:2015apa}, microlensing for Subaru (HSC) with dashed line shows the uncertain constraint of HSC and Eros/Macho \cite{Capela:2013yf,Niikura:2017zjd}, dynamical heating of ultra faint dwarf (UFD) \cite{Tisserand:2006zx}, CMB measurements \cite{Ali-Haimoud:2016mbv} and radio observation \cite{Gaggero:2016dpq}. 
 Taking into account these bounds, in Fig.~\ref{f9}  we superimpose our results for the PBH abundance using the parameters of the  first set in Table~\ref{tabu42}.
 This prediction is marked by  a black solid line reaching values for $\Omega_{PBH}/\Omega_{DM}$ up to $0.2$, between the  microlensing for Subaru
 and the white dwarfs explosion excluded regions.

Using  the last three  parameter sets   in Table~\ref{tabu4}, we superimpose our results in Fig.~\ref{f10}.  
Purple line corresponds to $\lambda / \mu=0.33327$, orange to $0.33330$ and green to $1/3$. 
%The initial conditions are given Table~\ref{tabu2} (second, third and fourth set respectively).
 We notice that,  although these three different parameter sets yield quite distinctive cosmological predictions, as can be seen in  Fig.~\ref{f6},
 by appropriate choice of the initial value for the field $\phi$ (see Table~\ref{tabu2}), we can achieve   almost 
  similar fractional abundance for all  cases,  as in Fig.~\ref{f10}. Finally, our results are consistent with the constraints calculated in ref.~\cite{Laha:2020ivk}.

\section{Conclusions}
\label{conclusion}

In this paper, we study  a model  based on a no-scale supergravity  with  $SU(2,1)/SU(2) \times U(1)$   symmetry~\cite{Ellis:2013xoa}, with  a  deformed  K\"ahler potential 
by introducing a simple exponential  term, using practically one extra parameter. The perturbation due to this modification  
 induces an inflection point to the effective scalar potential. As expected, this potential, in the absence of the modification yields the usual  Starobinsky-like  potential.
The superpotential  we employ is the well-known Wess-Zumino superpotential. 
 The induced inflection point   can be expounded as a peak  in the CMB power spectrum. 
 Interestingly enough,  using this mechanism we satisfy all the Planck cosmological constraints for inflation and we were able to achieve  ample 
 PBH production, that can explain up to 20\%--25\% of the DM of the Universe.

Moreover,  we studied the stability of the potential along the 
 inflationary directions, checking   all the  parameter sets presented in this work. 
Afterwards in the context of the 
slow-roll approximation, we use the single field inflation method  and we evaluate  the evolution of the field and the slow-roll parameters. 
We highlight that the slow-roll approximation fails to provide  the precise  power spectrum, therefore   the use  of the  
M-S equation is imperative. Eventually, using    the numerical result from the M-S solution, we calculate the  power spectrum and  the fractional abundance of PBH.

We have scanned  the parameters  entering in  the  modified  K\"ahler potential and we have     presented results for various sets of them.
Interestingly,   we have found  that potentials with   values for the  ratio $\lambda/\mu >  1/3$, which  are excluded by CMB constraints
 in the context of the original Wess-Zumino model, now
 become compatible with the latest Planck data. In parallel, 
     these values of the parameters are compatible to   significant amount of PBH.

 Unfortunately, as all the inflation models that use the inflection point mechanism in order to produce PBH, 
 our model requires fine-tuning of the parameters entering by the modification of the K\"ahler potential.
 Although, the numerical analysis reveals that this fine-tuning can be compensated in part,  by the 
 appropriate choice of the parameter $\delta_c$  that affects the calculation of the PBH abundance, a more 
 detailed quantitative analysis on this aspect can be performed. Moreover, 
 exploring inflationary models that are not using the inflection point mechanism in order to produce PBH,
 will alleviate the necessary fine-tuning. Both directions require detailed analysis, since the PBH is 
 an interesting alternative to the standard DM models.

\vspace{5mm}
\section*{Acknowledgments} 
%  \label{acknow}
The work of D.V.N was supported in part by the DOE grant DE-FG02-13ER42020 and in part by the Alexander S. Onassis Public Benefit Foundation.
The research work of V.C.S and  I.D.S. was supported by the Hellenic Foundation for Research
and Innovation (H.F.R.I.) under the “First Call for H.F.R.I. Research
Projects to support Faculty members and Researchers and the procurement of
high-cost research equipment grant” (Project Number: 824). V.C.S and I.D.S.  thank  I. Gialamas, A. Lahanas and N. Tetradis  for useful discussions. 
Finally,  I.D.S. would like to thank G. Ballesteros for useful communications.

\vspace{1cm}
%\clearpage

\end{document}